\documentclass[usenatbib,usegraphicx]{mn2e}

\newcommand{\chandra}{\textit{Chandra}}

\newcommand{\xspec}{{\sc xspec}}
\newcommand{\apec}{{\it apec}}
\newcommand{\fastwind}{{\sc fastwind}}
\newcommand{\zpup}{$\zeta$~Pup}

\newcommand{\taustar}{\ensuremath{\tau_{\ast}}}
\newcommand{\hinf}{\ensuremath{h_{\infty}}}
\newcommand{\Ro}{\ensuremath{{R_{\mathrm o}}}}
\newcommand{\Rcl}{\ensuremath{{R_{\mathrm cl}}}}
\newcommand{\fv}{\ensuremath{{f_{\mathrm v}}}}

\newcommand{\Rstar}{\ensuremath{{R_{\ast}}}}
\newcommand{\Lstar}{\ensuremath{{L_{\ast}}}}

\newcommand{\Rsun}{\ensuremath{\mathrm {R_{\sun}}}}
\newcommand{\Lsun}{\ensuremath{\mathrm {L_{\sun}}}}
\newcommand{\Msun}{\ensuremath{\mathrm {M_{\sun}}}}
\newcommand{\Msunyr}{\ensuremath{{\mathrm {M_{\sun}~{\mathrm yr^{-1}}}}}}
\newcommand{\hetgs}{HETGS}

\newcommand{\kms}{km s$^{-1}$}
\newcommand{\vinf}{\ensuremath{v_{\infty}}}
\newcommand{\Teff}{\ensuremath{T_{\rm eff}}}
\newcommand{\Lx}{\ensuremath{L_{\rm X}}}
\newcommand{\Lbol}{\ensuremath{L_{\rm Bol}}}
\newcommand{\Lya}{${\rm Ly}{\alpha}$}
\newcommand{\Ha}{${\rm H}{\alpha}$}

\newcommand{\Mdot}{\ensuremath{\rm \dot{M}}}
\newcommand{\NH}{\ensuremath{N_\mathrm{H}}}

\newcommand{\apj}{ApJ}
\newcommand{\apjs}{ApJS}

\newcommand{\aap}{A\&A}

\newcommand{\aj}{AJ}
\newcommand{\mnras}{MNRAS}
\newcommand{\araa}{ARAA}
\newcommand{\pasp}{PASP}
\newcommand{\pasa}{PASA}

\begin{document}

\title[HD~93129A \chandra\/ Spectroscopy]{{\it Chandra} X-ray
  spectroscopy of the very early O supergiant HD 93129A: constraints
  on wind shocks and the mass-loss rate}

\author[D. Cohen et al.]{David H.\ Cohen,$^{1}$\thanks{E-mail:
    cohen@astro.swarthmore.edu} Marc Gagn\'{e},$^{2}$ Maurice A.\
  Leutenegger,$^{3,4}$ James P.\ MacArthur,$^{1}$ \newauthor Emma E.\
  Wollman,$^{1,5}$ Jon O. Sundqvist,$^{6}$ Alex W.\ Fullerton,$^{7}$
  Stanley P.\ Owocki$^{6}$
  \\
  $^{1}$Swarthmore College, Department of Physics and Astronomy, Swarthmore, Pennsylvania 19081, USA\\
  $^{2}$West Chester University, Department of Geology and Astronomy, West Chester, Pennsylvania 19383, USA \\
  $^{3}$NASA/Goddard Space Flight Center, Code 662, Greenbelt, Maryland 20771, USA \\
  $^{4}$CRESST and University of Maryland, Baltimore County, MD 21250, USA \\
  $^{5}$Caltech, Department of Physics, 1200 East California Blvd., Pasadena, California 91125, USA \\
  $^{6}$University of Delaware, Bartol Research Institute, Newark,
  Delaware 19716, USA \\
  $^{7}$Space Telescope Science Institute, 3700 San Martin Dr., Baltimore, Maryland 21218, USA \\
}

\maketitle

\label{firstpage}

\begin{abstract}

  We present analysis of both the resolved X-ray emission line
  profiles and the broadband X-ray spectrum of the O2 If$^{\ast}$ star
  HD 93129A, measured with the \chandra\ HETGS. This star is among the
  earliest and most massive stars in the Galaxy, and provides a test
  of the embedded wind shock scenario in a very dense and powerful
  wind. A major new result is that continuum absorption by the dense
  wind is the primary cause of the hardness of the observed X-ray
  spectrum, while intrinsically hard emission from colliding wind
  shocks contributes less than 10\%\/ of the X-ray flux. We find
  results consistent with the predictions of numerical simulations of
  the line-driving instability, including line broadening indicating
  an onset radius of X-ray emission of several tenths \Rstar.
  Helium-like forbidden-to-intercombination line ratios are consistent
  with this onset radius, and inconsistent with being formed in a
  wind-collision interface with the star's closest visual companion at
  a distance of $~100$ AU. The broadband X-ray spectrum is fit with a
  dominant emission temperature of just k$T = 0.6$ keV along with
  significant wind absorption. The broadband wind absorption and the
  line profiles provide two independent measurements of the wind
  mass-loss rate: $\Mdot = 5.2_{-1.5}^{+1.8} \times 10^{-6}$ \Msunyr\/
  and $\Mdot = 6.8_{-2.2}^{+2.8} \times 10^{-6}$ \Msunyr,
  respectively.  This is the first consistent modeling of the X-ray
  line profile shapes and broadband X-ray spectral energy distribution
  in a massive star, and represents a reduction of a factor of 3 to 4
  compared to the standard \Ha\/ mass-loss rate that assumes a smooth
  wind.

\end{abstract}


\section{Introduction} \label{sec:intro}

With a spectral type of O2 If$^{\ast}$ \citep{Walborn2002}, HD 93129A
is among the earliest, hottest, most massive and luminous O stars in
the Galaxy. As such, it has an extremely powerful wind, with a
terminal velocity in excess of 3000 \kms, and a mass-loss rate thought
to be in excess of $10^{-5}$ \Msunyr\/ \citep{Repolust2004}. In fact,
HD 93129A has been considered to have the highest mass-loss rate of
any O star in the Galaxy \citep{Taresch1997,Benaglia2004}. This star,
therefore, provides an interesting test of the wind-shock paradigm of
massive star X-ray emission, both because of the tremendous kinetic
power in its wind and because of the dominant role played by X-ray
absorption in such a dense wind.  In order to study the X-ray emission
and absorption we have analyzed the \chandra\/ \hetgs\/ spectrum with
a focus on individual line profiles.  In addition, we analyze the
low-resolution, zeroth-order ACIS CCD spectrum in order to determine
the relative contributions of high-temperature thermal emission and
wind attenuation to the observed spectral hardness.  These
complementary analysis techniques provide information about the
temperature and kinematics of the X-ray emitting plasma, about its
spatial distribution, and about the wind mass-loss rate.

X-ray emission from O stars is attributed to three mechanisms: (1)
Embedded Wind Shocks (EWS), generally assumed to be associated with
the Line-Driving Instability (LDI)
\citep{lw1980,ocr1988,fpp1997,Kahn2001}; (2) Colliding Wind Shocks
(CWS) in some binary systems \citep{Stevens1992,Antokhin2004,pp2010};
and (3) Magnetically Confined Wind Shocks (MCWS) for stars with
significant dipole magnetic fields \citep{bm1997,uo2002,Gagne2005}. Of
these, the EWS mechanism is assumed to operate in all O stars, while
CWS may dominate in massive binaries with strong enough winds and MCWS
in those small number of O stars with strong, large-scale magnetic
fields. The EWS mechanism produces plasma of several million degrees
and is associated with relatively soft X-ray emission, while the other
two mechanisms produce stronger shocks, higher temperatures, and
harder X-ray emission.  However, it should be kept in mind that soft
X-ray absorption by the bulk wind can harden the observed X-rays from
EWS in O stars with high mass-loss rates \citep{Leutenegger2010}.

In high-resolution X-ray spectra, the hallmark of embedded wind shocks
is broad emission lines \citep{Kahn2001,Cassinelli2001}. By analyzing
the widths and profile shapes of individual X-ray emission lines in
the grating spectra of O stars, the kinematics of the hot, X-ray
emitting plasma embedded in the warm, partially ionized bulk wind can
be determined, testing the predictions of the EWS scenario.
Furthermore, due to preferential absorption of red-shifted line
photons from the far hemisphere of O star winds, X-ray emission lines
from embedded wind shocks have a characteristic blue-shifted and
skewed shape, in proportion to the characteristic wind optical depth
\citep{oc2001}.  It recently has been shown that a wind mass-loss rate
can be determined by fitting the ensemble of derived characteristic
optical depths, given a model of the bulk wind X-ray opacity
\citep{Cohen2010}. The initial application of this technique to
\zpup\/ (O4 If) provided a mass-loss rate determination that was
roughly a factor of three lower than the traditional value derived
from the strength of the \Ha\/ emission under the assumption of a
smooth, unclumped wind.  This lower value is consistent with other
recent reassessments \citep{Puls2006} that do account for small-scale
wind clumping,
which affects density-squared diagnostics such as \Ha\/ emission
strength.  The associated clumping factors are consistent with those
seen in numerical simulations of the LDI \citep{ro2002,do2005}.
Indeed, there is a consensus emerging that many, if not all, O stars'
mass-loss rates must be lowered by factors of several due to the
effects of clumping \citep{hfo2008}. In this context, the X-ray line
profile mass-loss rate diagnostic is especially useful, as it is not
affected by small-scale clumping, as long as the individual clumps are
optically thin to X-rays \citep{Cohen2010}.

Both numerical simulations and the lack of significant observed X-ray
variability indicate that clumps in O star winds are on quite small
scales, with sizes $\ell \ll \Rstar$. Since even the entire wind of HD
93129A is only marginally optically thick to bound-free absorption of
X-rays, it is very likely that such small clumps will be individually
quite optically thin to X-rays.  This means they cannot have much of
the self-shadowing that would reduce the exposure of wind material to
X-rays, and would thus lead to a significant ``porosity'' reduction in
the overall absorption. An important point here then is that, while
such porosity requires a strong clumping (with individual clumps that
are optically thick), a clumped wind need not be porous (if the clumps
are optically thin).\footnote{In this paper we refer to ``optically
  thin clumping'' when discussing clumps that do not have a porosity
  effect but do have an effect on \Ha, reserving ``porosity'' to
  describe the effects of optically thick clumps.  For the unmodified
  term ``clumping'', the reader should bear in mind that
  density-squared diagnostics such as \Ha\/ will still always be
  affected, but that X-ray transmission will be affected by an
  associated porosity if and only if the clumps are optically thick.
  Finally, we note that a given clump or clump distribution may be
  optically thin at one wavelength and thick at others. }

HD 93129A lies in the Trumpler 14 cluster in the Carina nebula, at a
distance of 2.3 kpc, but with a modest visual extinction of $A_{\rm V}
= 2.3$ \citep{Townsley2011,Gagne2011}. It has a visual companion, HD
93129B, at a separation of 2.7\arcsec, with a spectral type of O3.5,
and it also has a closer visual companion (HD 93129Ab) detected using
the HST Fine Guidance Sensor (FGS) at a separation of 0.053\arcsec\/
\citep{Nelan2004,Nelan2010}. This closer companion is also estimated
to have a spectral type of O3.5.  Non-thermal radio emission has been
detected from the system, presumably indicating the existence of
colliding wind shocks \citep{Benaglia2006}. This motivated the initial
CWS interpretation of low-resolution \chandra\ CCD spectral
measurements of the relatively hard X-rays from HD 93129A
\citep{Evans2003}, although given the large separation of components
Aa and Ab (over 100 AU $\approx 1000$ \Rstar), any X-ray emission
associated with the wind interaction zone should be relatively weak
and the overall emission is likely dominated by EWS emission arising
much closer to the photosphere of component Aa. \citet{Gagne2011} has
recently suggested that only 10 to 15 percent of the X-ray emission is
due to CWS X-rays.  Furthermore, given the high wind mass-loss rate of
HD 93129A, significant attenuation of the soft X-ray emission is
expected.  This paper represents an attempt to account for wind
absorption when analyzing the star's X-ray properties.

Detailed analysis and modeling of the optical and UV spectra of HD
93129A was presented by \citet{Taresch1997}. That work included an
analysis of the interstellar absorption features in the UV, which
yielded a hydrogen column density measurement of $\NH = 2.5 \times
10^{21}$ cm$^{-2}$ that is basically consistent with the observed
reddening and inferred extinction ($\NH = 3.7 \times 10^{21}$
cm$^{-2}$ is implied by the color excess). Important results from the
UV and optical spectral analysis include a very high wind mass-loss
rate and terminal velocity in addition to a bolometric luminosity in
excess of $2 \times 10^6$ \Lsun\/ and evidence for non-solar
abundances in line with CNO processing (a nitrogen abundance several
times solar and carbon and oxygen abundances a factor of a few lower
than solar). These authors estimate a zero-age main sequence mass of
120 \Msun\/ for HD 93129A. More recent analysis by
\citet{Repolust2004} finds a significantly lower effective temperature
and a modestly lower bolometric luminosity along with a higher wind
mass-loss rate.  The stellar parameters derived from these two studies
are summarized in Table \ref{tab:properties}. We note that the
presence of the close binary companion will affect the radius
determination, and related quantities.  Given that the companion is
about a magnitude dimmer than the primary, this effect will be small
\citep{Repolust2004}.


\begin{table}
\begin{minipage}{80mm}
  \caption{Stellar and wind parameters adopted from
    \citet{Taresch1997} and \citet{Repolust2004} (in parentheses). }
\begin{tabular}{cc}
  \hline
  parameter & value \\
  \hline
  Mass & (95 \Msun) \\
  \Teff & 52000 K (42500 K) \\
  \Rstar & 19.7 \Rsun\/ (22.5 \Rsun) \\
  log \Lstar & 6.4 \Lsun\/ (6.17 \Lsun) \\
 $v_{\rm rot}{\rm sin}i$ & (130 \kms) \\
  \vinf & 3200 \kms\/ (3200 \kms) \\
 $\beta$ & 0.7 (0.8)  \\
  \Mdot & $1.8 \times 10^{-5}$ \Msunyr\/ ($2.6 \times 10^{-5}$ \Msunyr) \\
  \hline
\end{tabular}
\label{tab:properties}
\end{minipage}
\end{table}  

We describe the \chandra\/ \hetgs\/ data in \S2. In \S3 we analyze the
line profiles seen in the grating spectrum and also fit global thermal
emission models to the low-resolution but higher signal-to-noise
zeroth order spectrum. We discuss the implications of these analyses
for the three mechanisms of O star X-ray emission in \S4, and
summarize our conclusions in \S5.

\section{The \chandra\/ data}
\label{sec:data}

The data we use in this paper were taken between 8 November 2005 and 5
December 2005 in seven separate pointings, with a total effective
exposure time of 137.7 ks. All observations employed the Advanced CCD
Imaging Spectrometer with the High Energy Transmission Grating
Spectrometer (ACIS-S/HETGS) \citep{Canizares2005}, providing dispersed
spectra in the MEG and HEG grating arrays, as well as a low-resolution
CCD spectrum from the zeroth-order image. The grating data have been
presented previously in \citet{Westbrook2008} and \citet{Walborn2008},
and in \citet{Nichols2011}.

\begin{figure}
\vspace{0.55in}
\includegraphics[angle=0,scale=0.44]{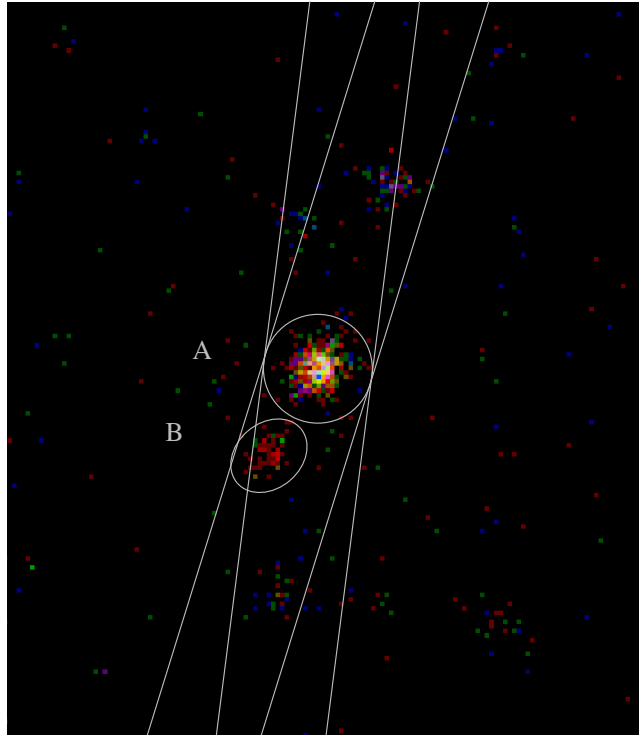}
\caption{The central region of the ACIS detector from the longest
  single exposure (Obs ID 7204; exposure time of 34 ks, corresponding
  to 25 percent of the total exposure time), showing the zeroth order
  image of HD 93129A (indicated by the circle), along with several
  other sources.  HD 93129B is labeled and located 2.7\arcsec\ to the
  southeast of component A (and indicated by the modestly elongated
  ellipse).  The detected photons are color coded according to energy,
  with low energies (0.5 to 1.5 keV) red, medium energies (1.5 to 2.5
  keV) green, and high energies (2.5 to 8 keV) blue.  Component B is
  clearly separable from component A and is softer and weaker (by
  roughly a factor of ten) than component A. Its relative weakness and
  softness are quantified in Fig.\ \ref{fig:zeroth_order}. The
  negative and positive MEG and HEG grating arm extraction regions are
  outlined in white (HEG is more vertical, MEG is angled to the upper
  right and lower left). Note that these have a somewhat different
  orientation in each observation. The zeroth order image of component
  B lies mostly within both of these extraction regions, indicating
  possible contamination of the dispersed spectra of HD 93129A. }
\label{fig:acis}
\end{figure}

The field is crowded with X-ray sources, but only HD 93129B, our
target's visual companion, is bright enough and close enough to HD
93129A to pose a potential problem in the data extraction.  In Fig.\
\ref{fig:acis} we show the center of the ACIS detector for the longest
of the seven separate exposures, with the MEG and HEG extraction
regions indicated, and the zeroth-order images of components A and B
labeled.  The CCDs that compose the ACIS detector have modest
intrinsic energy resolution (of roughly $E/{\Delta}E = 20$ to 50),
allowing us to assess the spectral energy distribution of each source
as indicated by the color coding in the figure.  For this particular
exposure, component B, which is 2.7\arcsec\ southeast of component A,
lies in the MEG and HEG extraction regions, and therefore is a
potential source of contamination of the dispersed spectra of
component A.


\begin{figure}
\includegraphics[angle=90,scale=0.37,width=85mm]{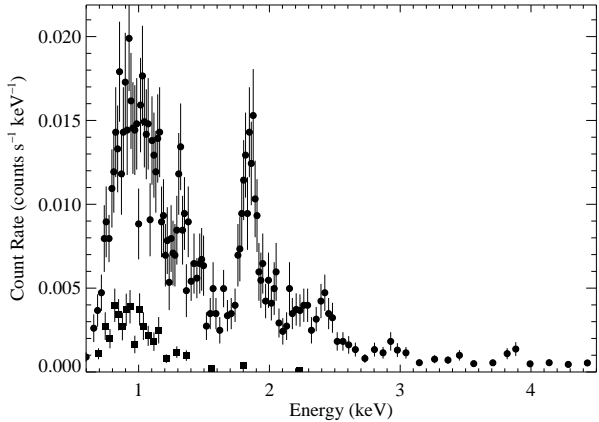}
\caption{The extracted ACIS CCD spectra -- zeroth order -- from the
  seven coadded pointings (circles) and the same for HD 93129B
  (squares).  Note that above 1.2 keV (below 10 \AA), the
  contamination of the grating spectra of component A by that of
  component B should be negligible.  }
\label{fig:zeroth_order}
\end{figure}

Due to the varying roll angle of the \chandra\/ instrument, component
B does not lie fully in the extraction regions of component A for all
of the seven separate observations.  Furthermore, as shown in Fig.\
\ref{fig:zeroth_order}, its photon flux is an order of magnitude lower
than that of component A.  Finally, the spectral energy distribution
of component B is much softer than that of component A (this can also
be seen, qualitatively, in the color coding of Fig.\ \ref{fig:acis}).
In fact, there are almost no counts from component B at energies
greater than 1.2 keV, corresponding to a wavelength of 10 \AA.  Thus,
the dispersed spectra of component A are unaffected by contamination
from component B below this wavelength. At wavelengths above 10 \AA,
however, there is likely to be modest contamination.  We note that the
2.7\arcsec\/ offset corresponds to a relative shift in the dispersed
MEG spectrum of roughly 1500 \kms, which could artificially broaden
the emission lines of component A.  We therefore restrict the analysis
of the grating spectra in this paper to wavelengths shortward of 10
\AA.


We note also that the close binary companion Ab detected at
0.05\arcsec with the FGS is completely unresolved by \chandra\/ so the
spectra we analyze in this paper are a composite of the two
components.  However, the spectral type of Ab is tentatively
identified as being the same as that of component B (O3.5 V)
\citep{Nelan2004}, so the composite spectrum, especially below 10 \AA,
is likely dominated by component Aa.  We discuss the colliding wind
X-ray emission due to this close visual companion in \S3.3.

\begin{figure*}
\includegraphics[angle=0,scale=0.22,width=170mm]{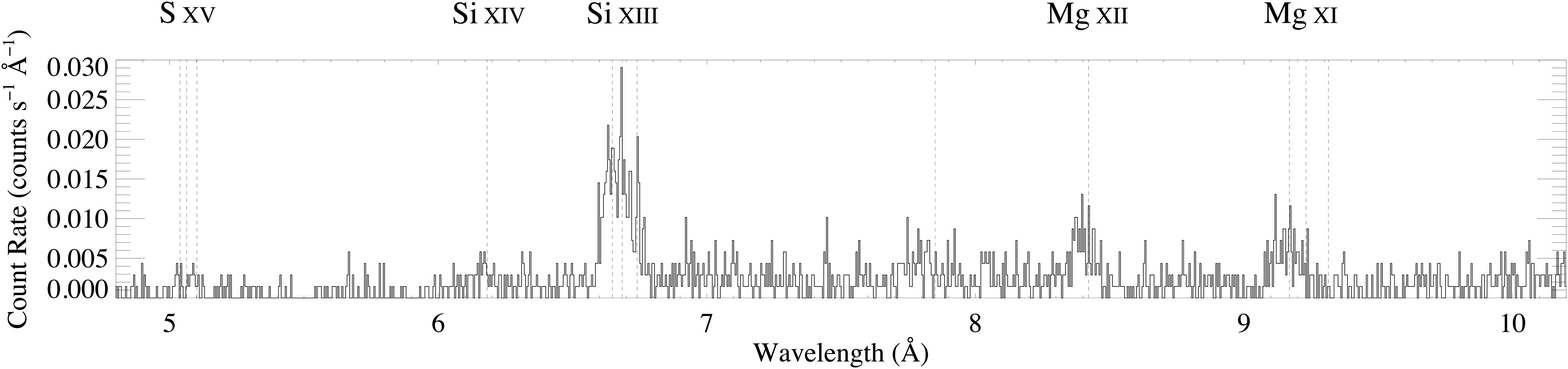}
\includegraphics[angle=0,scale=0.22,width=170mm]{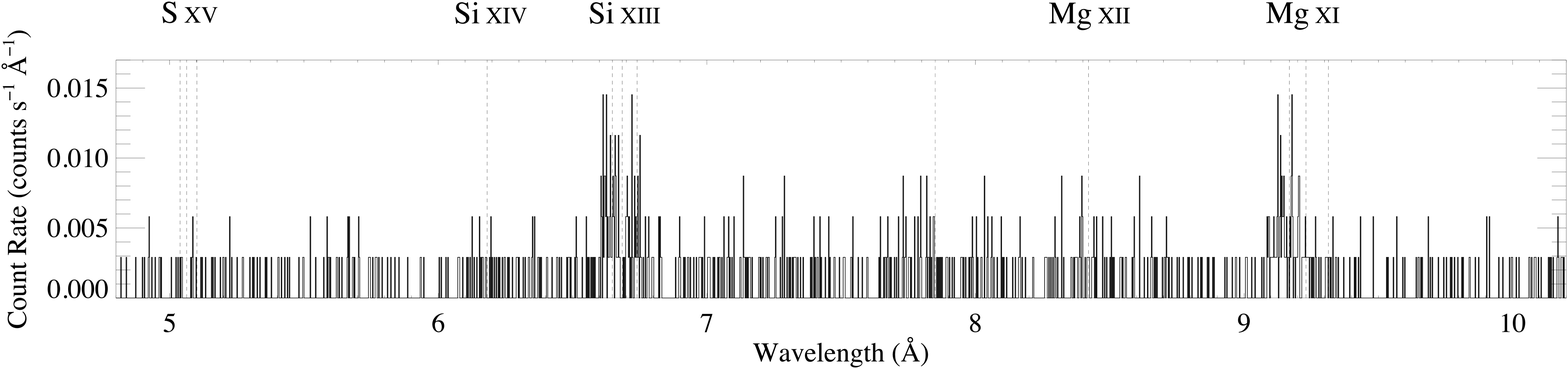}
\caption{The extracted MEG (top) and HEG (bottom) spectra from the
  seven coadded pointings. Note the different y-axis scales on the two
  figures. The wavelengths of lines expected to be present in normal O
  star \chandra\/ spectra are indicated by the vertical dotted lines.}
\label{fig:entire_spectrum}
\end{figure*}

After centroiding the zeroth-order image of component A in each of the
seven separate observations, we coadded the observations and extracted
the first-order MEG and HEG spectra, as well as the zeroth-order ACIS
CCD spectrum shown in Fig.\ \ref{fig:zeroth_order}.  The coadded
negative and positive first-order spectra (both MEG and HEG) are shown
in Fig.\ \ref{fig:entire_spectrum}. The MEG spectrum has a FWHM
resolution of 2.3 m\AA\/ and the HEG, with a lower sensitivity, has a
resolution of 1.2 m\AA.  In the 5 to 10 \AA\/ region where most of the
counts are, these correspond to resolving powers of roughly
$\lambda/{\Delta}\lambda$ = 300 to 600 in the MEG.

Even with the coaddition of seven separate pointings, for an effective
exposure time of 137.7 ks, the spectra have quite low signal-to-noise.
This is due both to the great distance to HD 93129 (2.3 kpc) and also
to the large interstellar column density, which causes significant
attenuation of the soft X-ray emission from the star. As we will show
in \S\ref{sec:analysis}, there is also significant X-ray attenuation
from the dense stellar wind.  The appearance of the \chandra\/ grating
spectra of HD 93129A is certainly significantly harder than that of
other normal O stars, but this is due primarily to attenuation and not
to high temperatures, as can be seen qualitatively in Fig.\
\ref{fig:entire_spectrum} from the dominance of the He-like lines
(Si\, {\sc xiii}, Mg\, {\sc xi}) over the corresponding H-like lines
(Si\, {\sc xiv}, Mg\, {\sc xii}), especially for silicon. This
qualitative impression is borne out by quantitative modeling presented
in \S3.3.

Only a handful of lines are present in the MEG spectrum (and even
fewer in the lower signal-to-noise HEG spectrum). The small number of
visible lines is affected by the overall low signal-to-noise and the
aforementioned attenuation, which renders the normally quite strong
Fe\, {\sc xvii} and O lines longward of 15 \AA\/ completely absent.
And it is exacerbated by the very large line widths, which spread the
modest number of line photons over many pixels.  After discarding the
very weak lines longward of 10 \AA\/ due to contamination from
component B, we are left with five lines and blended line complexes
detected with greater than 3$\sigma$ significance: the He-like S\,
{\sc xv} complex near 5.1 \AA, the \Lya\/ line of Si\, {\sc xiv} at
6.18 \AA, the He-like Si\, {\sc xiii} complex near 6.7 \AA, the Mg\,
{\sc xii} \Lya\/ line at 8.42 \AA, and the He-like Mg\, {\sc xi}
complex near 9.2 \AA. The weak Mg\, {\sc xi} He$\beta$ line at 7.85
\AA\/ is not detected above the 3$\sigma$ threshold. Not shown in the
figure are very weak Ne\, {\sc ix} and Ne\, {\sc x} lines between 12
\AA\/ and 14 \AA, which are affected by contamination by the softer
X-ray emission from component B, as discussed above.


\section{Spectral analysis}
\label{sec:analysis}

\subsection{Resolved emission lines}

By fitting a simple, empirical line profile model \citep{oc2001} to
the Doppler-broadened emission lines, we can simultaneously determine
the kinematics of the X-ray emitting plasma and the degree of
attenuation by the wind in which the hot, shock-heated plasma is
embedded. The specific parameters of the \citet{oc2001} model are the
onset radius of the X-ray emission (\Ro) and the fiducial optical
depth of the bulk wind, $\taustar \equiv
{\kappa}\Mdot/4{\pi}\Rstar\vinf$.  Note that \taustar\/ is expected to
vary from line to line due to the wavelength dependence of the wind
opacity, $\kappa$.

The fitted \Ro\/ values derived from the data are expected to be
several tenths of a stellar radius above the photosphere, based on
simulations of the line-driving instability (LDI)
\citep{ocr1988,fpp1997,ro2002}. By deriving values for this parameter
from the individual lines in the \chandra\/ spectrum of HD 93129A, we
can test the LDI scenario for embedded wind shocks in the most extreme
O star wind. And the values we derive for the fiducial optical depth,
\taustar, can be used to derive a mass-loss rate by fitting the
ensemble of values, given a model of the wind opacity, as has been
shown by \citet{Cohen2010}.

Following the procedure described in \citet{Cohen2010}, we assess the
continuum level near each line by fitting a small region of the
spectrum on either side of the line, and then fit a profile model plus
the continuum model (with the level fixed at the value found from
fitting the nearby continuum). We allow the normalization, \Ro, and
\taustar, to be free parameters of the fit, while fixing the velocity
law parameter at $\beta = 0.7$ \citep{Taresch1997} and the terminal
velocity at the value determined from the analysis of UV observations,
$\vinf = 3200$ \kms\/ \citep{Taresch1997, Repolust2004}.  We find the
best-fit model parameters by minimizing the C statistic
\citep{Cash1979}, and assign confidence limits individually to each
model parameter (while allowing the other free parameters to vary)
according to the formalism in \citet{Press2007}. We fit the MEG and
HEG data simultaneously.  We perform all of this modeling and data
analysis in {\sc xspec} v.12.6, using the custom model {\it
  windprofile}\footnote{The {\it windprofile} model's implementation
  in {\sc xspec} is described at
  heasarc.gsfc.nasa.gov/docs/xanadu/xspec/models/windprof.html, as is
  the {\it hewind} model we use to fit helium-like complexes.}.

For the helium-like complexes, which effectively comprise three
blended lines each (resonance (r), intercombination (i), and forbidden
(f)), we fit three profile models simultaneously (using the custom
model {\it hewind}), with the \Ro\/ and \taustar\/ parameters tied
together for each of the three lines, as described in
\citet{Leutenegger2006}.  The overall normalization and the
$\mathcal{G} \equiv (f+i)/r$ ratio are explicit fit parameters, while
the diagnostically important $\mathcal{R} \equiv f/i$ ratio is a
function of radius, via the radial dependence of the photoexcitation
rate of electrons out of the upper level of the forbidden line into
the upper level of the intercombination line.  This physics is
controlled in our modeling by the same \Ro\/ parameter that describes
the onset radius of X-ray emission. Lower values of \Ro\/ give more
plasma close to the photosphere, where it is more strongly affected by
photoexcitation, providing a lower $\mathcal{R} \equiv f/i$ value than
if \Ro\/ were larger and the overall photoexcitation rate were lower.
Thus, \Ro\/ controls both the line widths and the relative strengths
of the intercombination and forbidden lines in the observed spectrum
in a self-consistent manner. For the photoexcitation modeling, we use
UV fluxes from a TLUSTY $\Teff = 42500$ K, log $g = 3.75$ model
atmosphere \citep{lh2003}, and the atomic parameters from
\citet{Dere2007}.

The number of free parameters in the models -- both for the single
lines and for the He-like complexes -- is kept to a minimum.  We fix
the wind terminal velocity ($\vinf = 3200$ \kms) and the velocity law
parameter ($\beta = 0.7$) at the value determined from the UV data.
There is an extensive discussion of the sensitivity of the important
model parameters, \taustar\/ and \Ro, to these, and other, fixed
parameters as well as the effect of such factors as background
subtraction and continuum placement in \S4.3 of \citet{Cohen2010}, to
which we refer the reader.

\begin{figure}
\includegraphics[angle=90,scale=0.2,width=85mm]{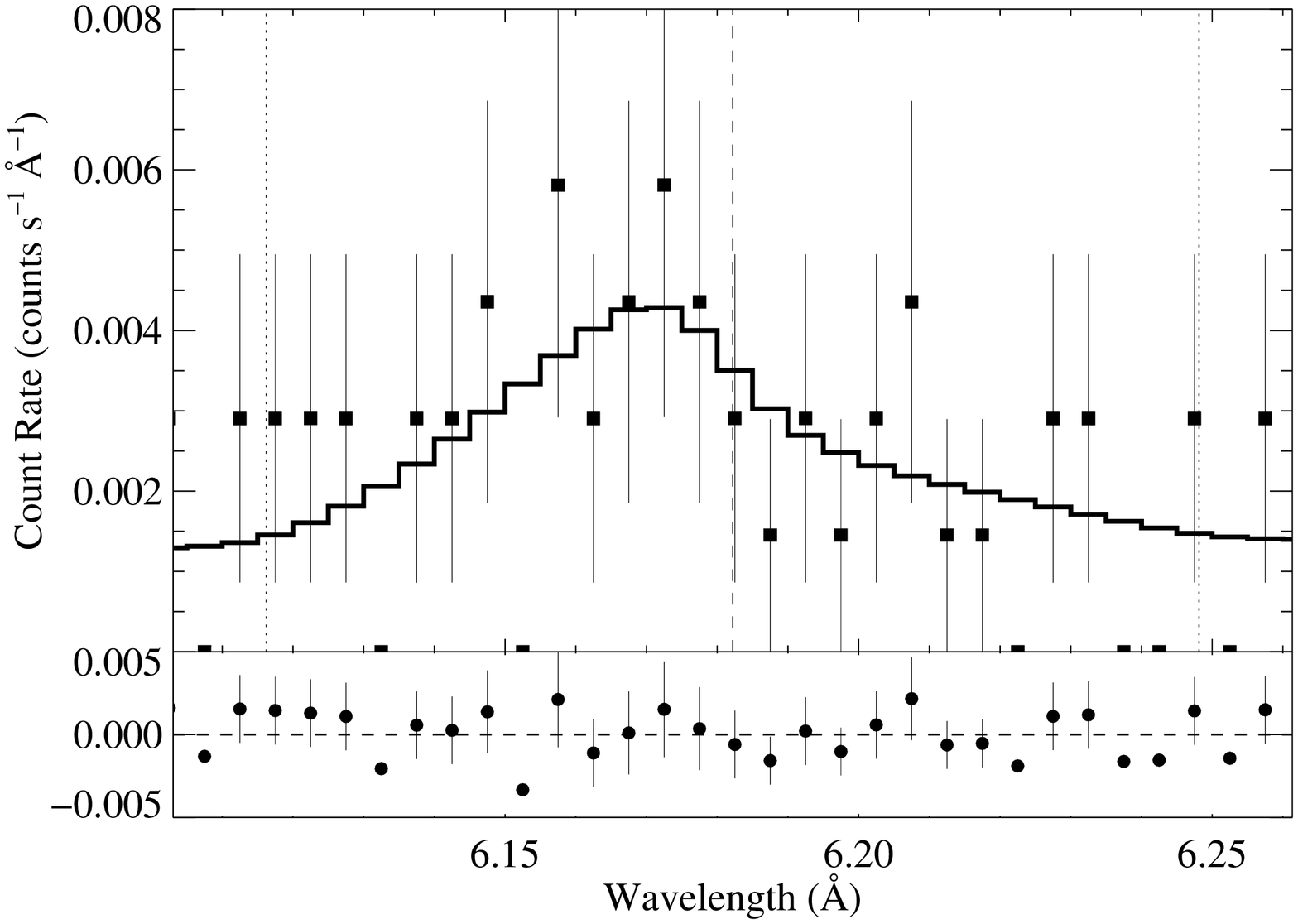}
\includegraphics[angle=90,scale=0.2,width=85mm]{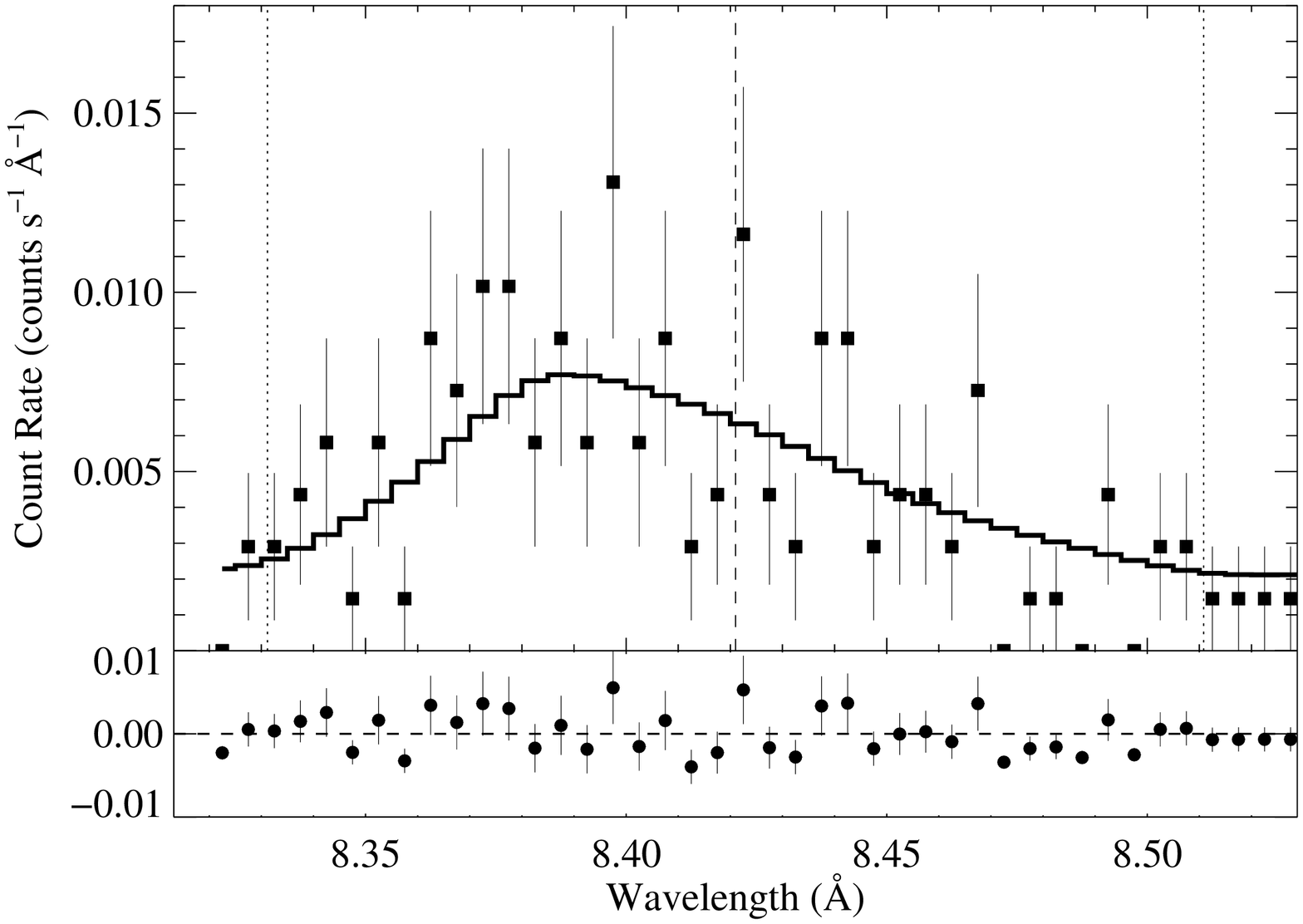}

\caption{The MEG data for both individual lines with sufficient
  signal-to-noise to warrant profile fitting, along with the best-fit
  profile model for each line (histogram). These are the
  Ly$\alpha$ lines of Si\, {\sc xiv}, at 6.18 \AA\/ (top), and Mg\,
  {\sc xii}, at 8.42 \AA\/ (bottom). The laboratory rest wavelength of
  each line is indicated by a vertical dashed line and the Doppler
  shifts associated with the (positive and negative) terminal velocity
  are indicated by the vertical dotted lines.  Poisson error bars are
  indicated on each data point. Note that the best-fit model for the
  Mg\, {\sc xii} line is based on jointly fitting the HEG and MEG
  data, though we show only the MEG data here.  }
\label{fig:individual_lines}
\end{figure}

\begin{figure}
\includegraphics[angle=90,scale=0.2,width=85mm]{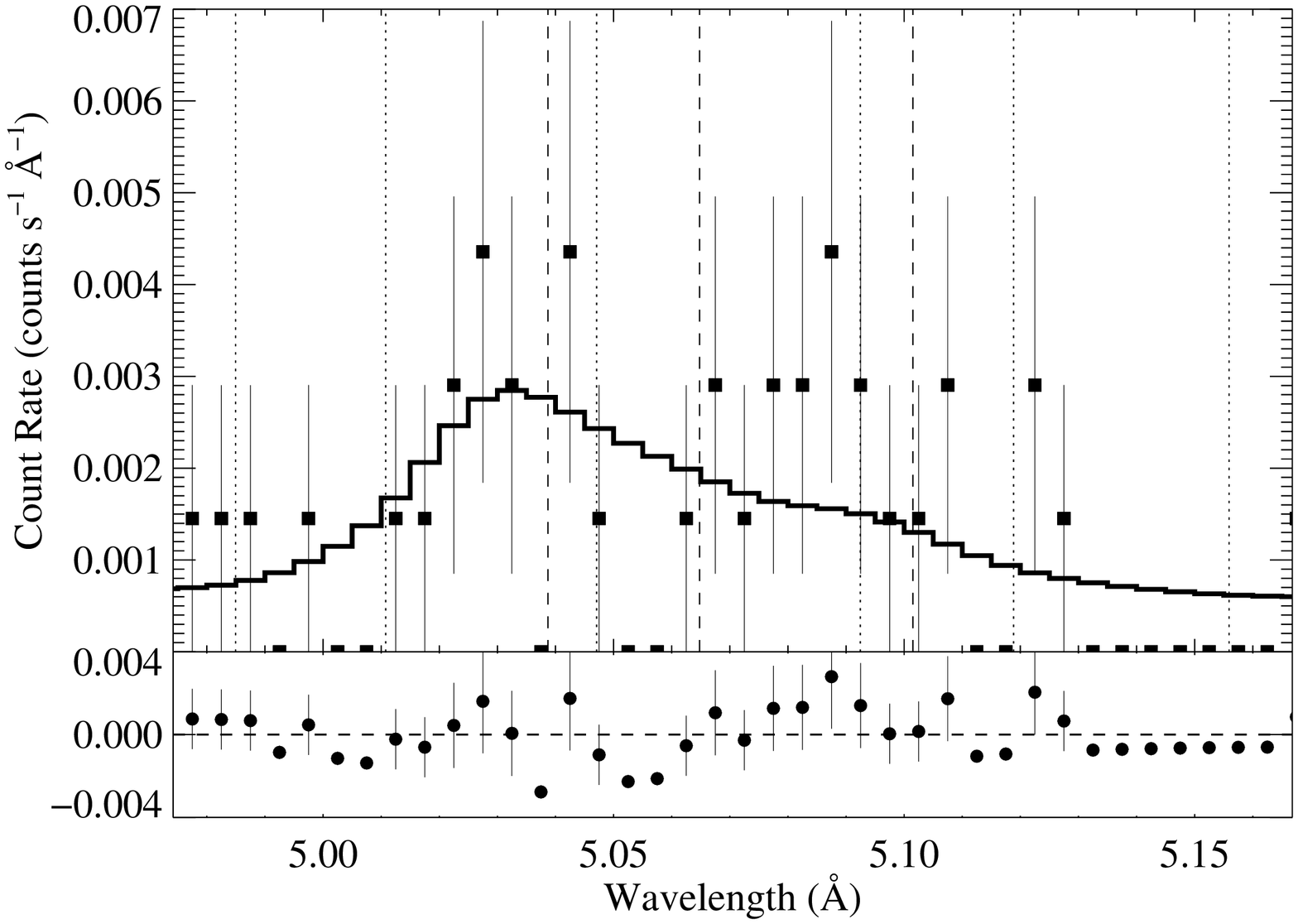}
\includegraphics[angle=90,scale=0.2,width=85mm]{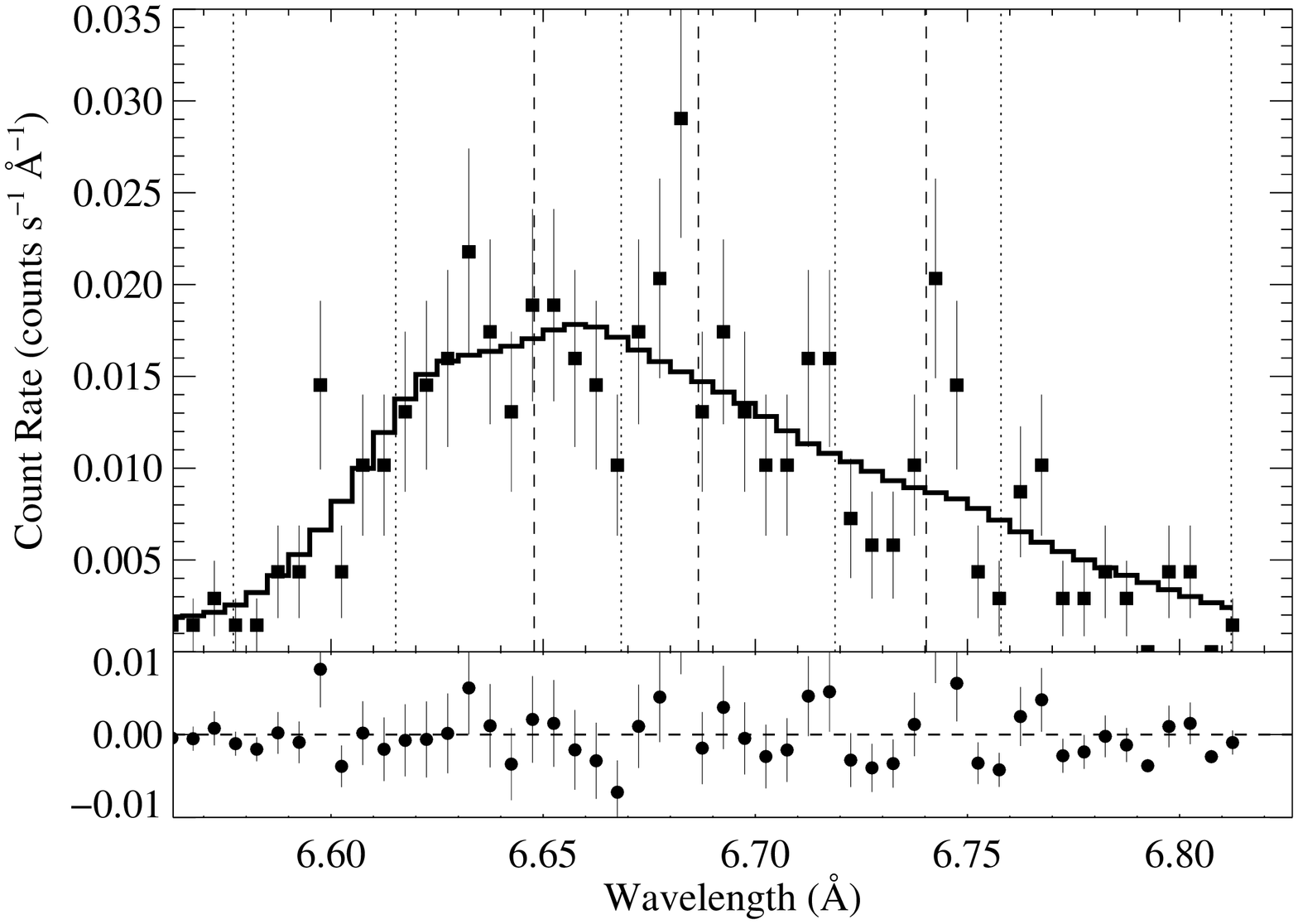}
\includegraphics[angle=90,scale=0.2,width=85mm]{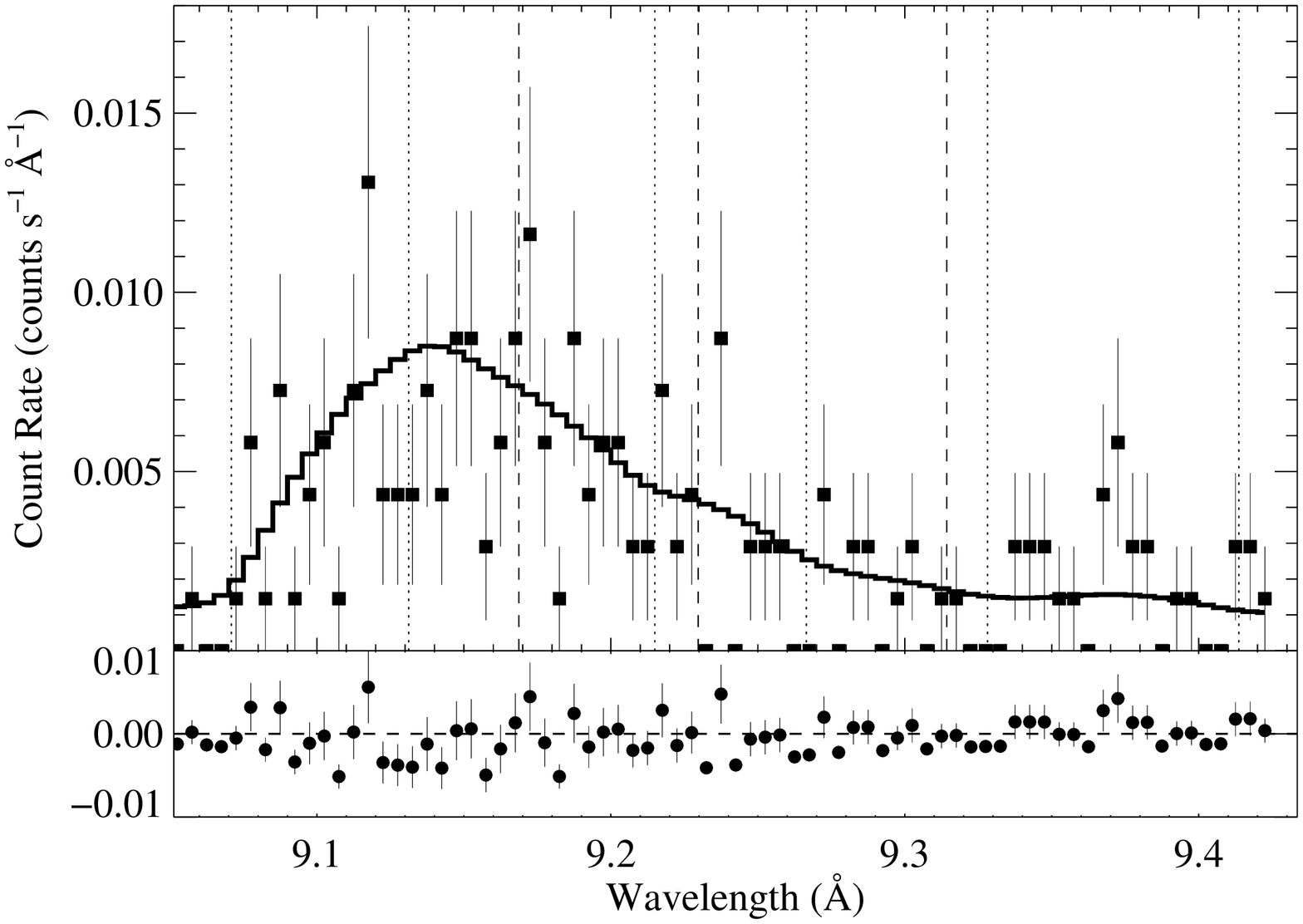}
\caption{The MEG data for the helium-like line complexes: S\, {\sc xv}
  (top), Si\, {\sc xiii} (middle) and Mg\, {\sc xi} (bottom). We show
  line center and terminal velocity indicators as in the previous
  figure, but here, we show three groups; one for each of the three
  lines in each complex.  The best-fit models are based on jointly
  fitting the HEG and MEG data (except for the weaker S\, {\sc xv}
  complex, where we fit only the MEG data). Note the very weak
  forbidden line in both the Mg and Si complexes (longest wavelength
  of the three lines in each complex). }
\label{fig:he-like_complexes}
\end{figure}

The results of our wind profile fits to the five lines and line
complexes are shown in Fig.\ \ref{fig:individual_lines} for the single
\Lya\/ lines and Fig.\ \ref{fig:he-like_complexes} for the three He-like
line complexes.  The quantitative results are summarized in Tab.\
\ref{tab:line_fits}. We note that when we increase the wind velocity
parameter, $\beta$, from 0.7 to 1.0, the characteristic optical depth
values, \taustar, increase by roughly 30\%, while the onset radii,
\Ro, increase by several tenths of a stellar radius. Similarly, there
is sensitivity of the important derived parameters to the assumed wind
terminal velocity.  For a higher terminal velocity of 3400 \kms, the
best-fit \taustar\ values decrease by roughly 20\%\/ and increase by
the same amount when we use a lower terminal velocity of 3000 \kms.
The onset radii, \Ro, vary by roughly 10\%\/ for these changes in
terminal velocity. We note that these systematic uncertainties --
especially for \taustar\/ -- are small compared to the statistical
errors.


\begin{table*}
\begin{minipage}{90mm}
  \caption{Wind profile model fit results}
\begin{tabular}{ccccc}
  \hline
  ion & wavelength\footnote{Closely spaced doublets in the \Lya\/ lines and He-like intercombination lines are fit with a single profile model centered at the emissivity-weighted wavelength of the two components.} & \taustar\ & \Ro & normalization\footnote{For the He-like complexes, the total normalization of all the lines in the complex is indicated.}  \\
  & (\AA)      &           & (\Rstar) & ($10^{-6}$ ph cm$^{-2}$ s$^{-1}$) \\  
  \hline
  S\, {\sc xv} & 5.0387, 5.0648, 5.1015 & $0.32_{-.32}^{+1.36}$ & $1.16_{-.15}^{+.44}$ & $4.11_{-1.11}^{+1.14}$  \\
  Si\, {\sc xiv} & 6.1822 & $1.64_{-1.26}^{+1.80}$ & $1.01_{-.01}^{+.52}$ & $1.48_
  {-.41}^{+.49}$  \\
  Si\, {\sc xiii} & 6.6479, 6.6866, 6.7403 & $0.81_{-.39}^{+.63}$ & $1.37_{-.14}^{+.12}$ & $16.8_{-0.8}^{+1.1}$  \\
  Mg\, {\sc xii} & 8.4210 & $1.03_{-.59}^{+1.12}$ & $1.44_{-.27}^{+.27}$ & $4.00_{-.47}^{+.64}$  \\
  Mg\, {\sc xi} & 9.1687, 9.2297, 9.3143 & $2.84_{-1.59}^{+.72}$ & $1.20_{-.19}^{+.88}$ & $16.1_{-1.4}^{+1.1}$  \\
  \hline
\end{tabular}
\label{tab:line_fits}
\end{minipage}
\end{table*}  

Before we move on to interpreting these model-fitting results, we
report on a few experiments involving fitting different types of
models to these same emission lines. Specifically, Gaussian profiles
have traditionally been fit to the broadened emission lines seen in O
stars. Therefore, we fit the highest signal-to-noise single line in
the spectrum, the Mg\, {\sc xii} \Lya\/ line, with an unshifted
Gaussian.  As expected from the asymmetry seen in the bottom panel of
Fig.\ \ref{fig:individual_lines}, the fit is poor (the wind profile
fit is preferred at $>$ 99.99\%). Allowing the Gaussian centroid to be
a free parameter, we find an improved fit with a large centroid blue
shift (to $\lambda = 8.401$ \AA), equivalent to $-730 \pm 200$ \kms,
and a width of $\sigma = 1200_{-170}^{+200}$ \kms.  This
shifted-Gaussian fit is statistically indistinguishable from the wind
profile model fit.  Although the {\it windprofile} fit is more
meaningful, as it is based on a physically realistic model while
having no more free parameters than the Gaussian model, the Gaussian
fit confirms and quantifies the large line widths and blue shifts
expected from EWS emission.

\begin{figure}
\includegraphics[angle=90,scale=0.2,width=85mm]{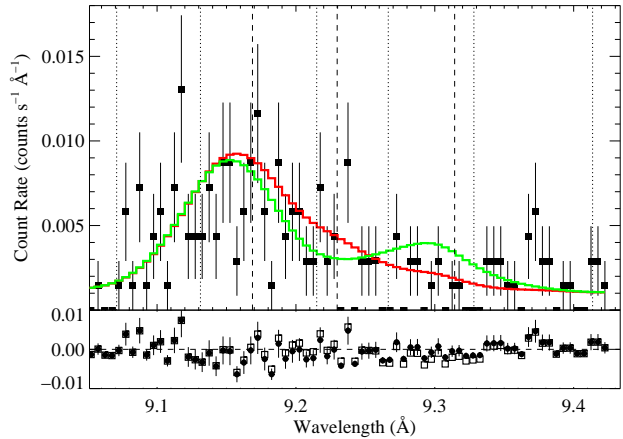}

\caption{The MEG data for the Mg\, {\sc xi} line complex fit with
  three shifted Gaussians.  When the $f/i$ ratio is a free parameter
  (red), it is low ($0.4_{-.17}^{+.23}$). When the ratio is fixed at
  the normal, equilibrium limit ($f/i = 2.7$) with no photoexcitation,
  as would be expected far from the star (green), the fit is poor. }
\label{fig:hegauss}
\end{figure}

Along similar lines, He-like complexes can be fit with three
Gaussians, providing a direct measure of the $f/i$ ratio, from which a
single radius of formation can be inferred.  We show these results in
Fig.\ \ref{fig:hegauss}. For the Mg\, {\sc xi} complex we find a low
ratio of $0.40_{-.17}^{+.23}$.  We next fit the same model, but with
the $f/i$ ratio fixed at the ``low density limit'' that would be
expected if the X-ray plasma were very far from the photosphere, where
photoexcitation cannot alter the ratio. This fit (which is also shown
in Fig.\ \ref{fig:hegauss}) is poor compared to the $f/i = 0.4$ fit,
implying that the hot plasma is relatively close to the photosphere.
This conclusion is in good agreement with the result from fitting the
three wind-profile model ({\it hewind}), in which the onset radius of
a distributed source of X-ray emitting plasma, \Ro, is constrained to
be within 1 stellar radius of the photosphere.


Porosity, due to optically thick clumps, can affect X-ray line
profiles by reducing the average optical depth of the wind
\citep{ofh2006}, and is another effect we might consider when fitting
the emission line data. However, the porosity length that is necessary
to provide a measurable effect is quite large \citep{oc2006} compared
to the small-scale structure in state-of-the-art 2-D radiation
hydrodynamics simulations \citep{do2003,do2005}. To explore the
effects of porosity on the fit quality and on the other parameters, we
fit the Mg\, {\sc xii} \Lya\/ line with a model having an effective
opacity modified by porosity from spherical clumps.  This model is
similar to the one described in \citet{oc2006} -- using the same
porosity-length ($h(r) \equiv \ell/f$, where $\ell$ is the
characteristic clump size and $f$ is the clump filling factor)
formalism -- but here employing a radial clump distribution determined
by the wind beta-velocity law, as used by \citet{ofh2006}. We fix the
characteristic optical depth at $\taustar = 4.2$, which is the value
we would expect at the wavelength of the Mg\, \Lya\/ line assuming a
mass-loss rate of $1.8 \times 10^{-5}$ \Msunyr\/ \citep{Taresch1997}.
The fit we obtain is similar in quality to the {\it windprofile} fit
reported in Table \ref{tab:line_fits}.  But in order to achieve this
good fit, a very large terminal porosity length of $\hinf = 5.3$
\Rstar\/ is required (with a 68\%\/ lower confidence limit of $\hinf =
2.5$ \Rstar), where the radially varying porosity length is given by
$h(r) = \hinf(1-\Rstar/r)^{\beta}$. If we use $2.6 \times 10^{-5}$
\Msunyr\/ as the standard, smooth-wind mass-loss rate
\citep{Repolust2004}, then the required porosity length is even
larger. Even the minimum porosity length of $\hinf = 2.5$ \Rstar\/ is
inconsistent with the numerical simulations of the line-driving
instability \citep{do2003,do2005}, requiring, for example, clumps that
are individually 0.25 \Rstar\/ in scale in a wind with a uniform
filling factor of $\fv = 0.1$.

\subsection{Mass-loss rate determination from the line profiles}

Given the characteristic optical depth values (\taustar, listed in
Table \ref{tab:line_fits}) obtained from fitting each line with the
{\it windprofile} model, we can derive constraints on the wind
mass-loss rate of HD 93129A. The characteristic optical depth is
defined as $\taustar \equiv \kappa\Mdot/4\pi\Rstar\vinf$, so that by
assuming a model of the bulk wind's X-ray opacity, $\kappa(\lambda)$,
as well as a value for the stellar radius and wind terminal velocity,
we can fit the wavelength-dependent \taustar\/ values with the
mass-loss rate as the only free parameter of the fit, as $\Mdot =
4{\pi}\Rstar\vinf\taustar(\lambda)/\kappa(\lambda)$. We have
demonstrated this process for analyzing the \chandra\/ grating
spectrum of the O supergiant \zpup\/ \citep{Cohen2010}. Because of the
lower quality of the \chandra\/ spectrum of HD 93129A compared to that
of \zpup, and because of the much smaller number (5 vs.\ 16) and
narrower wavelength range of usable lines, the wavelength trend in
\taustar\/ is not apparent for HD 93129A.  However, the five
\taustar\/ values are certainly consistent with the expected
wavelength trend.

We constructed a model of the bulk wind opacity for HD 93129A,
assuming that H and He are fully ionized, and assuming typical values
for the ionization balance of metals (generally dominated by triply
ionized states).  Furthermore, we assume solar abundances from
\citet{Asplund2009}, except for C, N, and O, which are altered by CNO
processing, according to the spectral analysis of \citet{Taresch1997}.
We take the C, N, and O values from these authors, but rescale them so
that the sum of the abundances of these three elements is equal to the
sum of the C, N, and O in the \citet{Asplund2009} solar abundances.
This effectively gives us $Z_{\rm C} = 0.25$, $Z_{\rm N} = 3.02$, and
$Z_{\rm O} = 0.47$, where \citet{Asplund2009} is the solar reference.
We should note, however, that because the emission lines we derive
\taustar\/ values from are all at short wavelengths, the alterations
to C, N, and O abundances have almost no effect on our results.
Changing the overall metallicity -- which we assume to be solar --
would have an effect, however. The derived mass-loss rate scales
inversely with the metallicity.


\begin{figure}
\includegraphics[angle=90,scale=0.37,width=85mm]{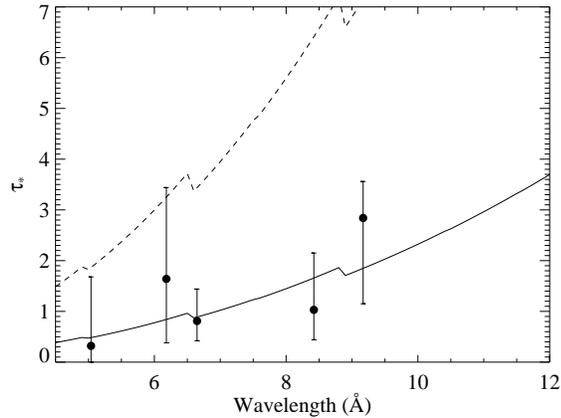}
\caption{The five \taustar\/ values fit with a single opacity model in
  order to derive \Mdot.  The best-fit \taustar\/ model ($\Mdot = 6.8
  \times 10^{-6}$ \Msunyr) is shown as a solid line, while the dotted
  line represents that form of \taustar\/ that would be expected if
  the traditional mass-loss rate of $2.6 \times 10^{-5}$ \Msunyr\/
  \citep{Repolust2004} were correct. }
\label{fig:taustar_mdot_fit}
\end{figure}

The result of fitting the \taustar\/ values for the mass-loss rate is
shown in Fig.\ \ref{fig:taustar_mdot_fit}. The best-fit mass-loss rate
is $6.8_{-2.2}^{+2.8} \times 10^{-6}$ \Msunyr, using the stellar
radius from \citet{Repolust2004}.  This represents a factor of three
\citep{Taresch1997} or four \citep{Repolust2004} reduction with
respect to traditional \Ha-based determinations that assume a smooth
wind (and thus ignore optically thin clumping).

\begin{figure}
\includegraphics[angle=90,scale=0.37,width=85mm]{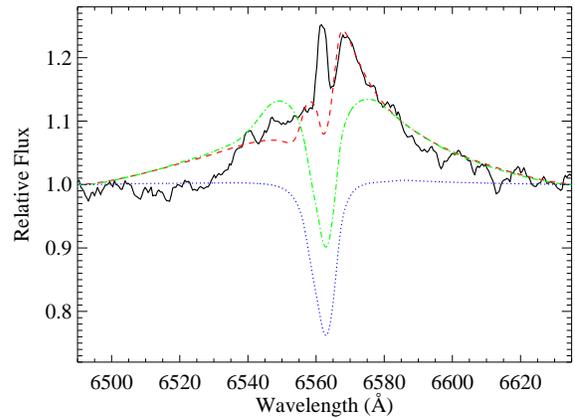}
\caption{The \Ha\/ emission profile (black, solid) is well reproduced
  by a model that includes clumping beginning at a radius of $\Rcl =
  1.05$ \Rstar\/ (red, dashed). The model with $\Rcl = 1.3$ \Rstar\/
  (green, dash-dot) does not have enough emission in the line core.
  For comparison, we show a model with no clumping (blue, dotted), in
  which the model fails to produce any emission. All models have
  $\Mdot = 7 \times 10^{-6}$ \Msunyr\/ and $\beta = 0.7$. Note that
  the narrow emission peak at line center likely has a significant
  contribution from nebular emission, which we do not model. }
\label{fig:halpha}
\end{figure}

To check for consistency with the observed \Ha, we modeled the profile
with the line-blanketed, non-LTE, unified (photosphere+wind) model
atmosphere code \fastwind\/ \citep{Puls2005}, which accounts for
optically thin clumping (using the filling factor approach) in the
calculations of the occupation numbers and the corresponding synthetic
spectra. Stellar parameters were taken from \citet{Repolust2004},
where the observational data also are described.  However, we adopt
the mass-loss rate derived from our X-ray analysis.  We show the data
in Fig.\ \ref{fig:halpha} along with three models, each with $\Mdot =
7 \times 10^{-6}$ \Msunyr\/ and $\beta = 0.7$. The best model has a
constant $\fv = 0.08$ above a radius $\Rcl = 1.05$ \Rstar, below which
the wind is assumed to be smooth. Note that this \Rcl\/ is not
necessarily the same as the X-ray onset radius, \Ro.  In fact, this
small clumping radius, \Rcl\/ is necessary to fit the data, as
comparison with the model that assumes $\Rcl = 1.3$ \Rstar\/ in Fig.\
\ref{fig:halpha} shows. In that model, the simulated strength of the
core is much too low, reflecting the reduced \Ha\/ opacity (which
scales as $\Mdot^2/\fv$) in the lower wind.  This early onset of wind
clumping has been found for a number of other stars as well
\citep{Bouret2005,Puls2006}. We also note that the simplest case of a
spatially constant \fv\/ actually reproduces the \Ha\/ data reasonably
well for this star, in contrast to, e.g., $\lambda$ Cep, for which a
rather strong radial dependence of \fv\/ is needed
\citep{Sundqvist2011}.


\subsection{Global thermal modeling}

While the individual line profiles provide information about the
kinematics of the shock-heated plasma, its spatial distribution, and
the degree of attenuation by the bulk wind in which the shock-heated
plasma is embedded, complementary information is contained in the
overall spectral energy distribution.  The broadband spectrum provides
information about the temperature distribution of the shock-heated
plasma and also the wind mass-loss rate via the degree of attenuation
of the X-rays. So, to complement the spectral modeling of individual
emission lines, described in the previous two subsections, we fit
global thermal models to the low-resolution zeroth-order spectrum of
component A shown in Fig.\ \ref{fig:zeroth_order}.

The model assumes optically thin, collisional-radiative equilibrium
(``coronal'') emission ({\it vapec}, the (Variable abundance)
Astrophysical Plasma Emission Code \citep{Smith2001}). It is
attenuated by the cold, neutral interstellar medium (via the \xspec\/
model {\it tbabs} \citep{Wilms2000}) and -- for the EWS emission --
the partially ionized stellar wind (via the \xspec\/ custom model {\it
  windtabs}\footnote{This model is also described on the {\it
    windprofile} custom model page on the \xspec\/ site.}
\citep{Leutenegger2010}). We model the EWS emission with a single
isothermal {\it vapec} spectrum. To account for the contribution of
harder X-rays from the wind-wind interaction we include a second
thermal emission component\footnote{Although the temperature
  distribution of the shocked plasma is certainly more complex than
  two discrete temperatures, we find that adding more temperature
  components does not improve the fit quality.} attenuated only by the
ISM.

This composite model is invoked in \xspec\/ as {\it
  (vapec$\times$windtabs + vapec)$\times$tbabs}.  The free parameters
of the model include the temperatures of the two {\it vapec}
components, their emission measures, the characteristic mass column,
$\Sigma_{\ast} \equiv \taustar/\kappa$ (g cm$^{-2}$), of the wind
absorption model, {\it windtabs}, and the interstellar column density.
Fixed parameters include the metallicity (fixed at solar, except for
CNO) of the emission model and the wind velocity profile (described by
$\beta = 0.7$) of the wind attenuation model.  We note that the {\it
  windtabs} model \citep{Leutenegger2010} has two features that make
it distinct from interstellar attenuation models and make it more
appropriate for the modeling attenuation by a stellar wind with
embedded shocks: (1) it incorporates atomic cross sections from
partially ionized species (e.g.\ O\, {\sc iv} rather than neutral O)
and assumes that H and He are fully ionized; and (2) it uses an exact
radiation transport model appropriate to an emitter spatially
distributed within the absorbing medium.  \citet{Leutenegger2010} show
that the attenuation from this realistic wind transport model differs
significantly from the exponential (``slab'') attenuation implemented
in ISM absorption models (their figures 5 and 8). For the fitting we
report on here, we use a solar abundance \citep{Asplund2009} wind
opacity model, but with altered C, N, and O abundances according to
\citet{Taresch1997}, just as we did for the analysis of the ensemble
of \taustar\/ values discussed in the previous subsection.  The {\it
  vapec} emission model assumes the same abundances as the {\it
  windtabs} absorption model.

We fit the above-described composite thermal emission with
wind-plus-ISM absorption model to the zeroth-order spectrum extracted
from the seven coadded pointings.  This low-resolution CCD spectrum
has significantly better signal-to-noise than the dispersed grating
spectra, but at a resolving power ($E/{\Delta}E$) of only a few tens.
It extends to lower energies (0.5 keV) than the dispersed spectrum
effectively does.  The zeroth-order spectrum does not suffer from
significant pile-up effects (unlike bare ACIS observations of the same
star). We use $\chi^2$ as the goodness of fit statistic and put
confidence limits on the fitted model parameters using the
$\Delta{\chi}^2$ formalism of \citet{Press2007}.

We let the interstellar column density be a free parameter of the fit,
and found a best-fit value of $\NH = 4.1 \times 10^{21}$
cm$^{-2}$, which is very close to the value implied by the color
excess ($\NH = 3.7 \times 10^{21}$ cm$^{-2}$ \citep{Gagne2011}).
The 68 percent confidence limits for $\NH {\rm (ISM)}$ extend
from $3.5 \times 10^{21}$ cm$^{-2}$ to $5.0 \times 10^{21}$ cm$^{-2}$.
The best-fit model has emission component temperatures of k$T = 0.61
\pm .02$ keV and k$T = 3.31_{-0.99}^{+2.11}$ keV, where the hotter
component has only 6 percent of the total emission measure. We
conjecture that this hotter component represents a small amount of CWS
X-ray emission, presumably associated with the non-thermal radio
emission detected in the system \citep{Benaglia2006}. Its contribution
is negligible below photon energies of 2.5 keV.  This provides further
confirmation that the wind-wind X-rays do not affect the line profiles
we discussed in the previous subsection.

\begin{figure}
\includegraphics[angle=0,scale=0.35,width=85mm]{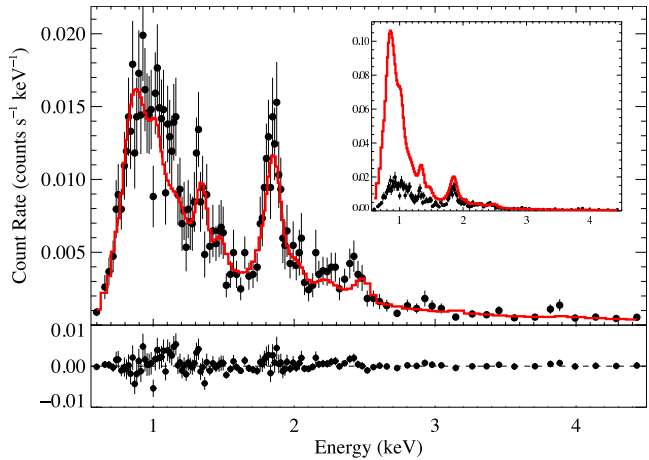}
\caption{The same zeroth order ACIS CCD spectrum shown in Fig.\
  \ref{fig:zeroth_order}, here fit with a two-temperature \apec\/
  thermal emission model (red histogram), where one temperature
  component (0.6 keV) is attenuated by the stellar wind as well as the
  interstellar medium and the other (3.3 keV) is attenuated only by
  the ISM. Note the presence of strong Si\, {\sc xiii} emission just
  below 2 keV.  The vast majority of the emission in this spectrum is
  line emission, but due to the low resolution of the detector as well
  as the presence of many weak, blended lines, the spectrum looks
  relatively smooth. The inset figure shows the same data with a model
  identical to the best-fit model, except that the wind absorption
  ($\Sigma_{\ast}$ in {\it windtabs}) is zeroed out.  This model
  spectrum makes the significance of the wind absorption effect quite
  obvious. Nearly 80\%\/ of the emitted EWS X-rays are absorbed before
  they can escape from the wind.  }
\label{fig:zeroth_order_fit}
\end{figure}

The characteristic wind mass column density, $\Sigma_{\ast}$ in the
{\it windtabs} model, is found to be $\Sigma_{\ast} =
0.0522_{-0.0146}^{+0.0185}$ g cm$^{-2}$. Using the wind terminal
velocity and stellar radius from \citet{Repolust2004}, this
corresponds to a mass-loss rate of $\Mdot = 5.18_{-1.45}^{+1.83}
\times 10^{6}$ \Msunyr.  The biggest contribution to the uncertainty
on the wind mass column density is the uncertainty on the interstellar
absorption\footnote{When we fix the ISM column density at $\NH = 2.5
  \times 10^{21}$ cm$^{-2}$ \citep{Taresch1997}, the wind mass column
  density, and along with it, the mass-loss rate, increases by a
  factor of 1.8.}.  The quality of the fit is good, with a reduced
$\chi^2$ of 1.01 for 113 degrees of freedom.  It is shown in Fig.\
\ref{fig:zeroth_order_fit}.

We note the consistency of the broadband mass-loss rate determination
with the independent determination from the ensemble of line profile
shapes, discussed in the previous subsection. The wind attenuation is
significant, as we show in Fig.\ \ref{fig:zeroth_order_fit}, where we
also include a model without wind attenuation.  The X-ray luminosity
(corrected for ISM attenuation) of the best-fit model is $L_{\rm X} =
7.29 \times 10^{32}$ ergs s$^{-1}$ (giving $\Lx/\Lbol = 1.3 \times
10^{-7}$), but when we also correct for wind attenuation, we find
$L_{\rm X} = 3.30 \times 10^{33}$ ergs s$^{-1}$.  This implies that 78
percent of the X-ray emission above 0.5 keV produced by embedded wind
shocks is absorbed before it escapes the wind (which can be seen
graphically in the inset of Fig.\ \ref{fig:zeroth_order_fit}). Note
that nearly all of the X-ray emission below 0.5 keV will also be
attenuated.


To further test the plasma emission temperature result, we can examine
the temperature-sensitive iron L-shell line complexes in the 10 to 17
\AA\/ wavelength range of the grating spectrometer.  The data do not
have good signal-to-noise in that wavelength region and likely also
suffer from modest contamination by dispersed photons from HD 93129B,
as we discussed in \S2. The contamination was primarily a concern with
respect to the emission line shapes. But a contamination level of 20
or 25 \%\/ (based on Fig.\ \ref{fig:zeroth_order}) will not
significantly skew the spectrum over a broad wavelength range. Because
we want to ignore the contaminated line shapes, and to enhance the
signal-to-noise ratio, we rebinned the MEG spectrum, using a 20 counts
per bin criterion. We then fit the 10 to 16 \AA\/
portion\footnote{Absorption renders the iron line complex near 17
  \AA\/ undetected.} of the MEG spectrum with an \apec\/ emission
model, including wind attenuation via {\it windtabs} and interstellar
absorption via {\it tbabs}. We fixed the wind mass column parameter of
{\it windtabs} at the best-fit value found in our global fitting
reported on earlier in this subsection and fixed the interstellar
column density at $\NH = 3.7 \times 10^{21}$ cm$^{-2}$.

Because of the large number of charge states with closely spaced
ionization energies, Fe L-shell emission is a sensitive diagnostic of
plasma temperature \citep{Behar2001}.  Ne-like Fe\, {\sc xvii}, with
strong emission lines near 15 and 17 \AA, is present over a moderately
wide range of temperatures, and dominates over higher charge states
below about k$T = 0.3$ keV. Fe\, {\sc xviii} has strong lines near
14.2 and 16 \AA\/ and Fe\, {\sc xx} has strong lines near 12.8 \AA.
Higher ionization states have strong lines primarily between 11 and 12
\AA. 

\begin{figure}
  \includegraphics[angle=0,scale=0.35,width=85mm]{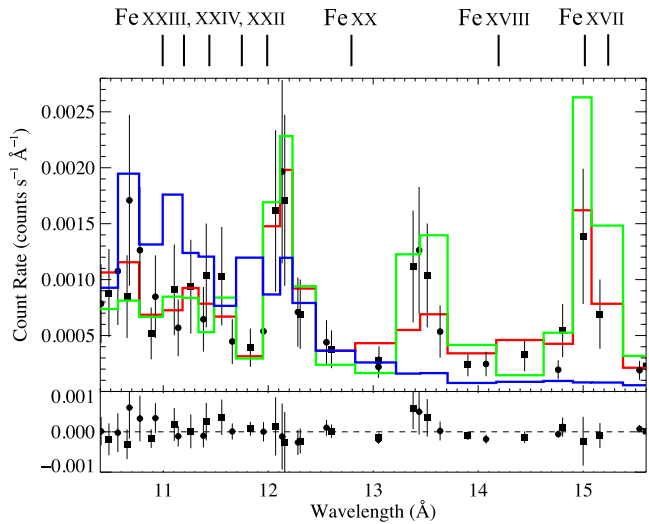}
  \caption{The binned MEG spectrum (squares are -1 order and circles
    +1 order) is shown along with the best-fit (MEG -1 order) model
    (red).  Cooler (0.3 keV, green) and hotter (1.5 keV, blue) models
    are also shown, making it clear that the presence of Fe\, {\sc
      xvii} emission near 15 \AA\/ in the data and the low levels of
    emission near 11 \AA\/ from higher ionization stages are
    incompatible with very hot (k$T > 1$ keV) plasma from colliding
    wind shocks. Ne\, {\sc ix} and {\sc x} emission complexes are also
    visible at 13.5 \AA\/ and 12.1 \AA, respectively.  Their ratio is
    temperature dependent and consistent with temperatures between 0.3
    and 0.6 keV. We indicate the wavelengths of some of the strongest
    lines at low and at high temperatures.  Note that there are other,
    low-temperature lines of, e.g., Ne\, {\sc ix} contributing to the
    observed flux in the short wavelength side of the spectral region
    displayed in this figure.}
\label{fig:FeL}
\end{figure}

In Fig.\ \ref{fig:FeL} we show the binned MEG data along with the
best-fit thermal emission model, which has a temperature of k$T =
0.58$ keV, in very good agreement with the zeroth order spectral
fitting.  Here, though, we can see which specific lines are and are
not contributing to the observed flux. The strong emission of Fe\,
{\sc xvii} is seen clearly near 15 \AA, but there is no strong
emission from the higher charge states.  Thus, the dominant plasma
temperature is constrained to be about 0.6 keV or lower, while only
very small contributions from higher temperatures are compatible with
the data.


\subsection{HD 93129B}

The zeroth order spectrum of HD 93129B, the O3.5 companion at a
separation of 2.7\arcsec, is shown in Fig.\ \ref{fig:zeroth_order}.
We have also fit it with a thermal emission model with both wind and
ISM absorption (but no CWS component), much as we did for HD 93129A.
Because of the low signal-to-noise of this spectrum, we held the
interstellar column density fixed at $\NH = 3.7 \times 10^{21}$
cm$^{-2}$, corresponding to the measured color excess.  Our best-fit
model has a temperature of k$T = 0.31 \pm .06$ keV and a wind
absorption mass column of $\Sigma_{\ast} = 0.0221_{-.0113}^{+.0267}$ g
cm$^{-2}$, corresponding to a mass-loss rate of $\Mdot \approx 1
\times 10^{-6}$ \Msunyr, with a factor of two uncertainty.

We note that this source emission temperature is somewhat lower than
that reported in \citet{Naze2011}.  This is most likely due to our
inclusion of wind absorption, which hardens the emergent spectrum.
The ISM-corrected X-ray luminosity of our best-fit model is $\Lx = 1.5
\times 10^{32}$ ergs s$^{-1}$, corresponding to $\Lx/\Lbol = 1.2
\times 10^{-7}$, very similar to the value we find for component A.

\section{Discussion}

The \chandra\/ observations of HD 93129A are consistent with the
embedded wind shock scenario that is generally assumed to apply to all
O stars. The X-ray luminosity is $\Lx = 1.3 \times 10^{-7} \Lbol$;
very much in line with the canonical value for O star wind-shock
sources \citep{Pallavicini1981}, although it is roughly a factor of
two higher than the average value found for O stars in Carina
\citep{Naze2011}. The emission temperature is quite low (0.6 keV), as
is expected from the EWS mechanism, although there is evidence for a
small ($\sim 6\%$) contribution from hotter plasma (3.3 keV). The
presence of this hotter component is not surprising, given the
detection of non-thermal radio emission from the system, presumably
associated with colliding wind shocks involving components Aa and Ab,
separated by roughly 100 AU.  But we stress that the colliding-wind
contribution to the overall X-ray spectrum is minimal.

Although the EWS emission temperature of 0.6 keV is low, the overall
spectrum is relatively hard.  We have shown here, using both the
individual line profiles and also the broadband, low-resolution CCD
spectrum, that this is due to attenuation by the star's dense stellar
wind. The two different, largely independent, manifestations of the
wind attenuation lead to consistent mass-loss rate determinations of
between 4.7 and $7.0 \times 10^{-6}$ \Msunyr, which represent a factor
of several reduction in the mass-loss rate over traditional values
determined from density-squared diagnostics.  As both X-ray mass-loss
rate diagnostics are insensitive to density squared effects, and
because some clumping is certainly expected in the wind of HD~93129A
\citep{Lepine2008}, the new, lower mass-loss rate seems quite
reasonable, and is in line with the factor of three mass-loss rate
reduction seen in \zpup\/ \citep{Puls2006,Cohen2010}. This is verified
by our modeling of the \Ha\/ line, which shows that $\Mdot = 7 \times
10^{-6}$ \Msunyr\/ provides a good fit if a constant clump volume
filling factor $\fv = 0.08$ is assumed. To explain the strong core of
the \Ha\/ line, a clump onset radius of $\Rcl = 1.05$ \Rstar\/ also
must be assumed.  We note that the X-rays are not produced at radii
this small. Presumably the wind shocks associated with clumps at the
slow-moving base of the wind are not strong enough to produce X-rays.

The kinematics of the X-ray emitting plasma in HD 93129A determined
from the line widths are consistent with the terminal velocity of the
bulk wind ($\vinf = 3200$ \kms), assuming a constant X-ray filling
factor above some onset radius, \Ro.  We derive \Ro\/ values for each
line or line complex from the profile fitting, and find onset radii
that are somewhat lower than, but still statistically consistent with,
the typical value of $\Ro \sim 1.5$ \Rstar\/ from numerical
simulations of the line-driving instability
\citep{ocr1988,fpp1997,ro2002}. However, if the wind terminal velocity
is lower than we assumed for our fitting, then the \Ro\/ values would
increase by 0.1 or 0.2, to $\Ro \sim 1.5$ \Rstar\/ (for an assumed
$\vinf = 3000$ \kms).

The location of the X-ray plasma is further constrained by the
forbidden-to-intercombination line ratios in the helium-like line
complexes observed in the grating spectra. Using a model that includes
both the broadening and attenuation effects on the line profiles in
conjunction with the altered forbidden-to-intercombination line ratios
due to UV photoexcitation, we find that all three helium-like line
complexes are completely consistent with the X-ray emitting plasma
being distributed throughout the wind of HD 93129A, starting at a
height of only several tenths of a stellar radius above the
photosphere.  Using a simpler model where we fit the
forbidden-to-intercombination line ratio directly, using Gaussian
profiles, we find low $f/i$ ratios for the Si\, {\sc xiii} and the
Mg\, {\sc xi} complexes.  For both these complexes, the ratio expected
if photoexcitation is unimportant, as would be the case in a binary
wind-wind collision zone far from either star's photosphere, is ruled
out with greater than 95\% confidence.

Finally, we must discuss alternative interpretations of the spectra
and, especially, the spectral lines. The profiles, while skewed and
blue-shifted, are not as asymmetric as expected given the very high
mass-loss rate traditionally found for the star.  Our interpretation
is that this is due to an actual mass-loss rate that is modestly lower
than the traditional value, but in principle, it could also be due to
porosity associated with optically thick and presumably large-scale
wind clumping. However, not only are the required porosity lengths
very high, but the same clumps that would need to be invoked to
generate a porosity effect would also lead to a mass-loss rate
reduction due to their effect on density-squared diagnostics.  If, for
example, the required porosity lengths were achieved with clumps
having a size scale of 0.25 \Rstar\/ and a uniform filling factor of
$f_v = 0.1$ in the context of an \Ha\/ mass-loss rate of $\Mdot = 1.8
\times 10^{-5}$ \Msunyr\/ that assumes no clumping
\citep{Taresch1997}, then the filling factor alone, via its effect on
the density-squared mass-loss rate, would reduce the mass-loss rate
inferred from the \Ha\/ to a value consistent with what we find from
the X-rays {\it without having to invoke any porosity} -- as we have
shown at the end of \S3.2 and in Fig.\ \ref{fig:halpha}.  Any
additional effect from porosity would make the mass-loss rates too low
and the X-ray line profiles too symmetric.

\section{Conclusions}

We have shown that the \chandra\/ grating spectrum of the extreme O
star, HD 93129A, can be understood using the same paradigm that
explains the canonical embedded wind shock source, \zpup, once
adjustments are made for its larger wind terminal velocity and
mass-loss rate.  Specifically, the kinematics of the X-ray emitting
plasma are consistent with shocks embedded in a $\beta = 0.7$, $\vinf
= 3200$ \kms\/ wind starting at several tenths \Rstar, and that the
attenuation signatures in the line profiles are consistent with a
mass-loss rate of 4.7 to $7.0 \times 10^{-6}$ \Msunyr, representing a
modest reduction compared to traditional mass-loss rates determined
from \Ha\/ measurements that ignore the effects of clumping, and
showing consistency with \Ha\/ modeling that includes modest clumping,
in line with what is seen in LDI simulations \citep{do2005}.  This
mass-loss rate reduction of a factor of three to four is consistent
with that found for \zpup\/ \citep{Cohen2010}. We have also
demonstrated for the first time that modeling wind absorption of
X-rays for both line profiles and for the broadband spectral energy
distribution leads to consistent results when a physically realistic
model of the broadband wind attenuation \citep{Leutenegger2010} is
used.

The global spectral modeling indicates that the dominant thermal
emission component has quite a modest temperature, of roughly 0.6 keV,
as predicted by EWS models.  The observed overall hardness of the
spectrum is attributable to wind attenuation, rather than high plasma
temperatures. The low dominant plasma temperature is also manifest in
the low Si\, {\sc xiv}/Si\, {\sc xiii} ratio, which is consistent with
the value found in the \chandra\/ grating spectrum of \zpup. There is
likely a small amount of hard X-ray emission from colliding wind
binary interaction between components Aa and Ab, many tens of AU from
either star's photosphere.  Because of the large separation of the
components, this X-ray emission makes a small contribution to the
overall X-ray spectral properties, representing less than 10\% of the
system's X-ray luminosity. The helium-like $f/i$ ratios also provide
evidence that the bulk of the X-rays arise in embedded wind shocks.

As HD 93129A is the earliest O star known, and has one of the
strongest winds of any O star, the work presented here strongly
suggests that the embedded wind shock scenario, as described by
numerical simulations of the line-driving instability, is widely
applicable to O stars, even those with extremely strong winds. And
that X-ray line profile analysis, especially in conjunction with
broadband spectral modeling, provides a good means of making a
clumping-independent mass-loss rate determination for O stars with
dense winds.

\section*{Acknowledgments}

Support for this work was provided by the National Aeronautics and
Space Administration through \chandra\/ award numbers AR7-8002X and
GO0-11002B to Swarthmore College.  EEW was supported by a Lotte
Lazarsfeld Bailyn Summer Research Fellowship and JPM was supported by
a Surdna Summer Research Fellowship, both from the Provost's Office at
Swarthmore College. MAL is supported by an appointment to the NASA
Postdoctoral Program at Goddard Space Flight Center, administered by
Oak Ridge Associated Universities through a contract with NASA. JOS
and SPO acknowledge support from NASA award ATP NNX11AC40G to the
University of Delaware. The authors thank V\'{e}ronique Petit for her
careful reading of the manuscript and several useful suggestions.






\end{document}